\documentclass[aps,prl,amsmath,amssymb,twocolumn,preprintnumbers,nofootinbib]{revtex4}
\usepackage[utf8]{inputenc} 
\usepackage{graphicx}
\graphicspath{{./figures/}}
\usepackage{url}
\usepackage[bookmarks, pagebackref=false]{hyperref}
\usepackage[usenames,dvipsnames]{xcolor}
\definecolor{orange}{cmyk}{0,0.5,1,0}
\definecolor{rossoCP3}{cmyk}{0,.88,.77,.40}
\definecolor{graa}{rgb}{0.8,0.8,0.8}
\definecolor{blaa}{rgb}{0.2,0.2,0.6}
	\hypersetup{
			colorlinks,
			bookmarksopen,
			bookmarksnumbered,
			citecolor=blaa, 		
			linkcolor=rossoCP3,	
			urlcolor=rossoCP3,			
			}
\usepackage{amsthm}
\usepackage{bm}
\usepackage{bbm}
\usepackage{pxfonts}

\usepackage{amsmath,amssymb,amsfonts}
\usepackage{color}
\usepackage{float}
\usepackage{hyperref}
\usepackage[Symbolsmallscale]{upgreek}
\usepackage{amsmath}
\usepackage{amsfonts}
\usepackage{amssymb,dsfont}
\usepackage{graphicx}
\usepackage{amssymb}
\usepackage[vcentermath]{youngtab}
\usepackage[all]{xy}
\usepackage{pstricks}
\usepackage{dsfont}%
\usepackage{bbold}

\usepackage{placeins}
\usepackage{xspace}
\usepackage{cancel} 

\usepackage{slashed}

\usepackage{natbib}

\usepackage{feynmf}

\usepackage{braket}

\newcommand{\beq}{\begin{eqnarray}}
\newcommand{\eeq}{\end{eqnarray}}

\newcommand{\bmp}{\noindent\begin{minipage}{16cm}}
\newcommand{\emp}{\end{minipage}\vskip 7mm} 

\newcommand{\Tr}{\text{Tr}}
\newcommand*{\del}{\mathop{\mathrm{{}\partial}}\mathopen{}}

\newcommand{\be}{\textbf{e}}

\def\lsim{\mathrel{\rlap{\lower4pt\hbox{\hskip1pt$\sim$}}
    \raise1pt\hbox{$<$}}}                
\def\gsim{\mathrel{\rlap{\lower4pt\hbox{\hskip1pt$\sim$}}
    \raise1pt\hbox{$>$}}}                

\begin{document}

\title{Charging the Walking U(N)$\times$U(N) Higgs Theory as a Complex CFT}
\author{Oleg {\sc Antipin}$^{\color{rossoCP3}{\clubsuit}}$}
\email{oantipin@irb.hr}
\author{Jahmall {\sc Bersini}
$^{\color{rossoCP3}{\clubsuit}}$}
\email{jbersini@irb.hr}
\author{Francesco {\sc Sannino} $^{\color{rossoCP3}{\diamondsuit},\color{rossoCP3}{\heartsuit}}$}
\email{sannino@cp3.dias.sdu.dk}
\author{Zhi-Wei Wang $^{\color{rossoCP3}{\diamondsuit}}$}
\email{wang@cp3.sdu.dk}
\author{Chen Zhang $^{\color{rossoCP3}{\spadesuit}}$}
\email{czhang@cts.nthu.edu.tw}
\affiliation{{ $^{\color{rossoCP3}{\clubsuit}}$ Rudjer Boskovic Institute, Division of Theoretical Physics, Bijeni\v cka 54, 10000 Zagreb, Croatia}\\{ $^{\color{rossoCP3}{\diamondsuit}}$\color{rossoCP3} {CP}$^{ \bf 3}${-Origins}} \& the Danish Institute for Advanced Study {\color{rossoCP3}\rm{Danish IAS}},  University of Southern Denmark, Campusvej 55, DK-5230 Odense M, Denmark. \\\mbox{ $^{\color{rossoCP3}{\heartsuit}}$Dipartimento di Fisica “E. Pancini”, Università di Napoli Federico II | INFN sezione di Napoli}\\ \mbox{Complesso Universitario di Monte S. Angelo Edificio 6, via Cintia, 80126 Napoli, Italy.}\\
{$^{\color{rossoCP3}{\spadesuit}}$ Physics Division, National Center for Theoretical Sciences, Hsinchu, Taiwan 300}}

\begin{abstract}
{We apply a semi-classical method to compute the conformal field theory (CFT) data for the U(N)xU(N) non-abelian Higgs theory in four minus epsilon dimensions at its complex fixed point. The theory features more than one coupling and walking dynamics.
Given our charge configuration, we identify a family of corresponding operators and compute their scaling dimensions which remarkably agree with available results from conventional perturbation theory validating the use of the state-operator correspondence for a complex CFT. }
  \\~\\
{\footnotesize  \it Preprint: RBI-ThPhys-2020-22, CP$^3$-Origins-2020-10 DNRF90}

\end{abstract}

\maketitle

\small
\section{Introduction}
Conformal field theory (CFT) tools play a central role in unveiling the dynamics of quantum field theories (QFT) in regimes where ordinary (non)perturbative methods are either inadequate or cumbersome. The approach is to investigate the dynamics of a desired class of QFT theories by considering certain CFT limits in their parameter space. Additionally, as we shall argue below, certain theory sectors  can be  directly investigated  using semiclassical approximations.

Here we consider the dynamics of the non-abelian $U(N)\times U(N)$ Higgs theory because it is ubiquitous in the literature due to its relevance to particle physics and cosmology. This theory features two marginal couplings in four dimensions that flow to zero at large distances and it develops a Landau pole at high energies making it an effective field theory. Removing the Landau pole is possible only if the couplings vanish along the entire renormalisation group. Theories such as this one are known as trivial QFTs. One can, however, embed the current theory into an asymptotically safe gauge-fermion-scalar theory like the one constructed in \cite{Litim:2014uca}. Here the theory is well defined at short distances because of the emergence of an interacting CFT. 

In this work we study the theory in four minus epsilon space time dimensions. We first show that in the infrared the model features two complex conjugated fixed points in the quartic couplings (complex CFTs) \cite{Paterson:1980fc}  signaling the appearance of a controllable near-conformal behaviour   of the \emph{walking} type \cite{Gorbenko:2018ncu,Gorbenko:2018dtm,Kaplan:2009kr,Benini:2019dfy,Faedo:2019nxw}.  The latter behaviour has been invoked in the literature for models of dynamical electroweak symmetry breaking~\cite{Holdom:1981rm,Holdom:1988gs,Holdom:1988gr}. Lattice methods have been employed to establish walking as summarised in Ref~\cite{Cacciapaglia:2020kgq}. Another way to discuss walking is via the emergence of two complex zeros of the beta-function in the near-conformal phase~\cite{Gorbenko:2018ncu,Sannino:2012wy}.  
 
We then employ a semiclassical approach to determine the scaling dimensions of a family of fixed charged operators to the leading  and next to leading order terms in the  charge expansion but to all-orders in the couplings.  Similar investigations appeared earlier in the literature corresponding to abelian models such as the $U(1)$ scalar $\phi^4$ theory  \cite{Badel:2019oxl,Arias-Tamargo:2019xld} and non-abelian examples such as the $O(N)$ model  \cite{Antipin:2020abu} in four minus epsilon dimensions,  the $O(N)$ model in six minus epsilon dimensions was studied in \cite{Arias-Tamargo:2020fow}. 

The novelties with respect to earlier results
 are that here we deal with non-abelian theories with more than one coupling and that the CFT is complex with an emerging walking behaviour.
{For a specific charge configuration we identify a family of corresponding operators and compute their scaling dimensions. Remarkably, we find agreement with available results stemming from conventional perturbation theory validating employing the state-operator correspondence~\cite{Cardy:1984rp,Cardy:1985lth} at a complex CFT. }  Our findings complement the non-perturbative large-charge approach employed in \cite{Orlando:2019skh,Orlando:2020yii} to investigate near-conformal dynamics.

 The large-charge limit was introduced recently in~\cite{Hellerman:2015nra,Alvarez-Gaume:2019biu} with the first non-supersymmetric four-dimensional application to the gauged-Yukawa version of this model at large charge presented in \cite{Orlando:2019hte}.

The work is organised as follows. We first introduce the theory and the associated complex CFT, we then introduce the charge configurations, construct the spectrum of operators at fixed charge and then analyse the low energy spectrum of the theory associated with the related charge configuration. Using the state-operator correspondence at the complex CFT we determine the ground state energy and associated scaling dimensions in the charge expansion. 
Then, combining the semiclassical approach with ordinary perturbation theory, we reconstruct the full two loops scaling dimension for the whole family of fixed-charge operators.
Remarkably, we are able to write it in a form valid for real and complex couplings, 
encompassing the case where both couplings are purely reals and the theory \emph {walks}.
Finally, we compare our choice of charge configuration to the one considered in \cite{Orlando:2019hte} and show its importance. 

\section{The $U(N)\times U(N)$ Higgs model as complex CFT }\label{SU(N)}
In Euclidean spacetime, the $U(N) \times U(N)$ linear sigma model is defined via the bare Lagrangian: \begin{equation} \label{model}
  \begin{aligned}
    \mathcal{L} =
   \Tr(\del_\mu H^\dagger \del^\mu H ) + u_0\Tr(H^\dagger H)^2 + v_0(\Tr H^\dagger H )^2
  \end{aligned}
\end{equation}
where  $H$ is a $N \times N$ matrix with complex entries, which transforms in the $(N, \bar{N})$ representation of the $U(N)\times U(N)$ symmetry and can be written in terms of $2N^2$ real scalar fields:
\beq
(H)_{a\alpha}=
\frac{\phi+i\eta}{\sqrt{2N_F}}\delta_{a\alpha}+\sum_{A=1}^{N^2-1} (h^A+i\pi^A)T^A_{a \alpha}
\label{matrixH}
\eeq
where $T^A_{a \alpha}$ are the generalized Gell-Mann matrices, normalized as $\rm{Tr} \left(T^A T^B\right) = \frac{1}{2} \delta_{AB}$.
The $N$-rescaled couplings of the model  are
$ \alpha_h  = \frac{u N}{(4\pi)^2} $ and $\alpha_v  = \frac{v N^2}{(4\pi)^2}$
and, in terms of these rescaled couplings, the one-loop beta functions in $4-\epsilon$ dimensions read
 \begin{eqnarray}
\beta_{\alpha_v}&=&-\epsilon \alpha_v+4\alpha_v^2\left(1+\frac{4}{N^2}\right) + 16\alpha_v\alpha_h + 12\alpha_h^2 
\\
\beta_{\alpha_h}&=&{}-\epsilon \alpha_h + \frac{24}{N^2}\alpha_v\alpha_h+8\alpha_h^2 
\ .
\end{eqnarray}
The beta functions to five loops have been derived in \cite{Calabrese:2004uk}.
 For $\alpha_h=0$, Lagrangian (\ref{model}) reduces to the $O(2 N^2)$ model with the 1-loop fixed point (FP) $\alpha_v^{O(2N^2)}=\frac{\epsilon N^2}{4(4+N^2)}$ while for $\alpha_h \neq 0$, two fixed points at 1-loop emerge and they are:
\begin{eqnarray} \label{FPS}
\alpha_v^* & = & \epsilon N^2 \frac{9+N\left(-N\pm i\sqrt{2N^2-6}\right)}{8(27-8 N^2+N^4)} \nonumber \\  \alpha_h^*&= &  -\epsilon N \frac{5N-N^3\pm 3i\sqrt{2 N^2-6}}{8(27-8 N^2+N^4)} \ .
\end{eqnarray}
For $N>\sqrt{3}$ they are complex thus defining two complex interacting CFT's.

 To  elucidate the impact of the complex CFT on the dynamics of the model we consider the infinite $N$ limit. Here the single-trace beta function $\beta_{\alpha_h}$ decouples from the double-trace one. An interacting fixed point in the infrared occurs for $\alpha_h^\ast = \epsilon/8$. Substituting this value in the beta function for the double-trace operator one  notice that the double-trace beta function is positive and has a minimum near the origin  controlled by $\epsilon^2$. Therefore  the running of $\alpha_v$ slows near this point, i.e. its running behaviour is replaced by a walking one. One can also show that such behaviour persists at finite $N$.

As we showed in \cite{Antipin:2020abu}, in the $O(N)$ model fixed-charge operators with the lowest scaling dimension are $Q$-index traceless symmetric tensors with classical dimension $Q$. For $Q=2$ and $N > 2$ (when representation \eqref{matrixH} is irreducible) we obtain the decomposition of the 2-index traceless symmetric tensor as
\begin{align}
&& {\small \Yvcentermath1  \yng(2)}_{\ O(2 N^2)} = (1,\text{Adj}) \oplus(\text{Adj},1)\oplus(\text{Adj},\text{Adj}) \nonumber  \\ &&  \oplus  \big[(\ {\tiny \Yvcentermath1  \yng(2)}\ ,{\tiny \Yvcentermath1  \yng(2)}\ ^* \ ) \oplus (\ {\tiny \Yvcentermath1  \yng(1,1)}\ ,{\tiny \Yvcentermath1  \yng(1,1)}\ ^* \ ) \oplus c.c. \big] \ .
    \end{align}
We computed the 1-loop scaling dimensions $\Delta_{\text{1-loop}}=2+\gamma_{\text{1-loop}}$ for the operators appearing in this decomposition, which are shown in the table below (with the scaling dimensions for the representations (1,\text{Adj}) and (\text{Adj},1) being identical):
\begin{equation}\begin{array}{c c c c c} \hline
{\rm Rep.} \  & (\text{Adj},1) & (\text{Adj},\text{Adj})& (\ {\tiny \Yvcentermath1  \yng(2)}\ ,{\tiny \Yvcentermath1  \yng(2)}\ ^* \ )  & \ (\ {\tiny \Yvcentermath1  \yng(1,1)}\ ,{\tiny \Yvcentermath1  \yng(1,1)}\ ^*)\\ \hline
\text{Oper.} & \rm{Tr}\left[H T^a H^{\dagger} \right] &  \rm{Tr}\left[T^a H T^a H^{\dagger} \right] & \rm{Tr}\left[K^i H \bar K^i H \right] &\rm{Tr}\left[L^i H \bar L^i H \right]\\ \hline
{\gamma_{\text{1-loop}}} & \frac{4 \alpha_v}{N^2} + 4 \alpha_h &  \frac{4 \alpha_v}{N^2} &  \frac{4\alpha_v}{N^2}+\frac{ 4 \alpha_h}{N} & \frac{4 \alpha_v}{N^2}-\frac{ 4 \alpha_h}{N}\\
\hline\end{array}%
\label{table}
\end{equation}%
where $K^i ( \bar K^i )$ and $L^i(\bar L^i)$ are the Clebsch-Gordan coefficients for the $SU(N)$ representations $\ {\tiny \Yvcentermath1  \yng(2)}\ (\ {\tiny \Yvcentermath1  \yng(2)}\ ^*  )$ and $ \ \ {\tiny \Yvcentermath1  \yng(1,1)}\ (\ {\tiny \Yvcentermath1  \yng(1,1)}\ ^*)$ respectively.
 The $\rm{Tr}\left[T^a H T^a H^{\dagger} \right]$ appeared earlier in \cite{Antipin:2014mga} and will be used to test our semiclassical computation. 
 For $\alpha_h=0$ the operators have the same scaling dimensions due to the enhanced $O(2N^2)$ symmetry.



\section{Charging the system} \label{Charge}

It would be desirable if we can probe the complex CFT associated with the complex
fixed points by methods that go beyond the conventional perturbation theory. In this regard the semiclassical
approach delineated in ~\cite{Badel:2019oxl} is well suited, which extracts the scaling dimensions
of the lowest-lying fixed-charge operators by virtue of the state operator correspondence~\cite{Hellerman:2015nra}.
We therefore perform a Weyl map to a cylinder of radius $R$ (i.e. $\mathbb{R}^d \to \mathbb{R} \times S^{d-1}$), with the cylinder action given by
\begin{align}
        \mathcal{S}_{cyl} = \int d^d x \sqrt{g} \ & 
%
\Big[ \Tr(\del_\mu H^\dagger \del^\mu H ) + u_0\Tr(H^\dagger H)^2 + v_0(\Tr H^\dagger H )^2
   \nonumber \\ &+ m^2 \Tr( H^\dagger H) \Big] \,.
\end{align}
Here $g$ denotes the metric determinant and $m^2=\left(\frac{d-2}{2R}\right)^2$
is the coefficient of the conformal coupling required by Weyl invariance. An operator with scaling dimension $\Delta$
in the CFT corresponds to a state with energy $E=\Delta/R$, which can be computed from a fixed-charge path integral
on the cylinder~\cite{Badel:2019oxl}. 

Following ~\cite{Orlando:2019hte}, it is simplest to consider a homogeneous ground state with the ansatz
(with $\tau$ being the cylinder time)
\begin{equation}
H_0\left(\tau\right)=e^{2iM \tau}B\ ,
\label{eq:ansatz}
\end{equation}
where $M$ and $B$ are diagonal matrices. For such solutions, the value of Cartan charges for the $SU(N)_L\times
SU(N)_R$ symmetries are encoded in the traceless charge matrices
\begin{align}
\mathcal{Q}_L\equiv -V\dot{H_0}H_0^\dagger=-2iVMB^\dagger B,\quad\mathcal{Q}_R\equiv VH_0^\dagger\dot{H_0}=2iVMB^\dagger B
\label{eq:noether}
\end{align}
with $V= R^{d-1} \Omega_{d-1}$ being the volume of $S^{d-1}$. Note $\mathcal{Q}_L+\mathcal{Q}_R=0$ follows as a consequence of our
diagonal ansatz solution Eq.~\eqref{eq:ansatz}. The normalization of the charge matrices is chosen such that
the operator $\Tr(\be_{21} H\be_{12} H^\dagger)$, with the $N\times N$ constant matrix $\be_{pq}$ defined by
$(\be_{pq})_{jk}=\delta_{jp}\delta_{kq}$, corresponds to the charge configuration
\begin{align}
\mathcal{Q}_{L,1/2}=-\mathcal{Q}_{R,1/2}=\rm{diag}\left\{-1/2,1/2,0,\cdots,0\right\} \ .
\label{eq:cf1}
\end{align}
A general fixed-charge operator with charge configuration $\mathcal{Q}=\mathcal{Q}_L=-\mathcal{Q}_R$ can be constructed
as, for example
\begin{align}
\Tr\Big[ \Pi_j (\tau_j H\tau_j^T H^\dagger)^{|y_j|}\Big] \ .
\label{eq:ocon2}
\end{align}
Here $y_j$ is an integer, determined by first choosing
a root basis ${\beta_j},j=1,2,...,N-1$, onto which we decompose $\mathcal{Q}$ as
\begin{align}
\mathcal{Q}=\Sigma_{j=1}^{j=N-1}y_j \hat{h}_{\beta_j} \ , 
\label{eq:qdec2}
\end{align}
where $\hat{h}_{\beta_j}$'s are the roots mapped into the Cartan subalgebra of $SU(N)$ and satisfy the normalization
condition $\Tr(\hat{h}_{\beta_j}^2)=\frac{1}{2}$, $\tau_j=\be_{pq}$ for some $p,q$ that depend on $j$, and should be
chosen to be an element in the root subspace of $\beta_j$. For a given charge configuration, one may
obtain variations of Eq.~\eqref{eq:ocon2} by changing the root basis, redistributing the trace operation, or
changing the order of different $\tau_j H\tau_j^T H^\dagger$ factors, giving rise to multiple operators corresponding to
a given charge configuration. In any case, the operator construction involves a decomposition equation like
Eq.~\eqref{eq:qdec2}. The normalization condition dictates entries of $\hat{h}_{\beta_j}$'s must be integers or half-integers,
and since $y_j$'s must be integers (see Eq.~\eqref{eq:ocon2}), we conclude entries of $\mathcal{Q}$ must be integers
or half-integers as well.

The operator identification for a generic charge configuration is complicated and not unique. However, it can be shown that
a special family of charge configurations
\begin{align}
\mathcal{Q}_{L,J}=-\mathcal{Q}_{R,J}=\rm{diag}\left\{-J,J,0,\cdots,0\right\} \ ,
\label{eq:cfqj}
\end{align}
with $J$ being a positive integer or half-integer, corresponds to a unique fixed-charge operator
$\mathcal{O}_J=\Tr[(\be_{21} H\be_{12} H^\dagger)^{2J}]$ with minimal classical scaling dimension $Q=4J$ living in the representation $(\Gamma_J,\Gamma_J)$ of $SU(N)_L\times SU(N)_R$, with the irreducible representation $\Gamma_J$ of $SU(N)$ defined through
its Dynkin label $(2J,0,\cdots,0,2J)$. For $J=1/2$, $\Gamma_J$ reduces to the adjoint of $SU(N)$. The uniqueness is related to the fact that this
special family of charge configurations corresponds to highest weights in tensor product of the adjoint representations.
More detail about operator identification from charge configurations is presented in appendix.

Motivated by the form of $\mathcal{Q}_{L,J},\mathcal{Q}_{R,J}$, we parameterize the $M,B$ matrices as
$M=-i\ \rm{diag}\left\{\mu,-\mu,0,\cdots,0\right\},B=\rm{diag}\left\{b,b,0,\cdots,0\right\}$, with $\mu>0,b>0$.
Then according to Eq.~\eqref{eq:noether} we have $J=2V\mu b^2$. The ansatz Eq.~\eqref{eq:ansatz} with this form
of $M,B$ satisfy the EOMs derived from the cylinder action as long as
\begin{align}
2\mu^2=(u_0+2v_0)b^2+\frac{m^2}{2} \ ,
\label{EOM}
\end{align}
which fixes $\mu$ and $b$ for given $J$ and $m$.

The fixed-charge path integral is equivalent to an unconstrained path integral with an effective action $\mathcal{S}_{eff}$
obtained by adding appropriate boundary terms  resulting in adding $16 \mu^ 2 b^2 $ to  $\mathcal{S}_{cyl}$~\cite{Badel:2019oxl} . The action $\mathcal{S}_{eff}$ evaluated on
the solution Eq.~\eqref{eq:ansatz} gives classical energy. To compute the leading quantum correction, we expand
$\mathcal{S}_{eff}$ around the fixed-charge solution Eq.~\eqref{eq:ansatz} to obtain an effective Lagrangian $\mathcal{L}_{quad}$
to quadratic order in the fluctuation field. The leading quantum correction is then computed through the functional
determinant associated with $\mathcal{L}_{quad}$, which is equivalent to a sum of dispersion relations over all degrees
of freedom.

The dispersion relations and their multiplicity can be explicitly worked out
\begin{equation}
\label{disp}
\begin{split}
\omega_1=&\sqrt{J^2_\ell+4\mu^2}\qquad 4N-8~\rm{d.o.f.}\\
\omega_2=&\sqrt{J^2_\ell+4(1-a_0)\mu^2+a_0 m^2}\qquad 2\left(N-2\right)^2\,\rm{d.o.f}\\
\omega_{3,4}=&\sqrt{J^2_\ell+4\mu^2}\mp 2\mu \qquad 2N-3~\rm{d.o.f. \ each}\\
\omega_{5,6}=&\sqrt{J^2_\ell+4\mu^2+2\left(4\mu^2-m^2\right)a_0}\pm2\mu\qquad \rm{one}~\rm{d.o.f.\,each}\\
\omega_{7,8}=&\sqrt{J^2_\ell+12 \mu ^2-m^2\pm\sqrt{16 \mu ^2p^2+\left(m^2-12 \mu ^2\right)^2}}\\
\omega_{9,10}=&\left(J^2_\ell+\left(8+4a_0\right)\mu ^2-a_0 m^2\right.\\
&\left.\pm\sqrt{\left[(8+4a_0)\mu ^2-a_0 m^2\right]^2+16 \mu ^2J^2_\ell}\,\right)^{\frac{1}{2}}\,,
\end{split}
\end{equation}
where $a_0=\frac{u_0}{u_0+2v_0}$, and $J^2_\ell = \ell (\ell+d-2)/R^2$ corresponds to the eigenvalues of the Laplacian on $S^{d-1}$.
Goldstone modes appear as a consequence of symmetry breaking enforced by fixing the charge. The symmetry breaking pattern
can be written as $\mathcal{G}_0\overset{\text{exp.}}{\longrightarrow}\mathcal{G}_1\overset{\text{spont.}}{\longrightarrow}\mathcal{G}_2$,
with
\begin{align}
\mathcal{G}_0 & \equiv SU(N)_L\times SU(N)_R\times U(1)_A \\
\mathcal{G}_1 & \equiv SU(N-2)_L\times SU(N)_R\times U(1)_{L3}\times U(1)_{L5}\times U(1)_A \nonumber\\
\mathcal{G}_2 & \equiv SU(N-2)_L\times SU(N-2)_R\times U(1)_{D3}\times U(1)_{D5}\times U(1)_{A6} \ . \nonumber
\end{align}
Here $SU(N-2)_L$ denotes the left $SU(N-2)$ transformations with generators living in the lower
$(N-2)\times (N-2)$ block, and the remaining $U(1)$'s are generated from:
\begin{align}
U(1)_{L3(D3)} & \rightarrow\rm{diag}\{1,-1,0,\cdots,0\}_{L(D)} \\
U(1)_{L5(D5)} & \rightarrow\rm{diag}\{1,1,0,\cdots,0\}_{L(D)} \\
U(1)_{A6} & \rightarrow\rm{diag}\{0,0,1,\cdots,1\}_A \ ,
\end{align}
where subscript L means left transformation, D means the diagonal part of
left and right transformation (i.e. $U(1)_{L+R}$), and finally A means axial part
of the left and right transformation (i.e. $U(1)_{L-R}$).
We omitted the extra vectorial $U(1)$ symmetry in the breaking pattern analysis above because it acts as a spectator. 

Since Lorentz symmetry is explicitly
broken, we expect both relativistic (Type I) and nonrelativistic (Type II) Goldstones~\cite{Nielsen:1975hm}. In fact the minus sign solution in $\omega_{3,4}$ corresponds to $2N-3$ Type II
Goldstones, while two additional Type I Goldstones come from the minus sign solution in $\omega_{7,8}$ and $\omega_{9,10}$, with
the one in $\omega_{7,8}$ being the conformal mode. When each Type I Goldstone is counted once and Type II Goldstone twice,
the sum of Goldstone degrees of freedom is $4N-4$ which matches the number of broken generators at the spontaneous breaking step,
saturating the Nielsen-Chadha bound~\cite{Nielsen:1975hm}.

\section{Semiclassical analysis and results} \label{computation}

The leading order (LO) ground state energy $E_{LO}$ is obtained by evaluating the effective action $\mathcal{S}_{eff}$ on the classical trajectory \eqref{eq:ansatz} and reads
\begin{equation}
\label{energy1}
    E_{LO}=  \frac{ m N \left(48 \left(\frac{\mu}{m}\right)^4 - 8 \left(\frac{\mu}{m}\right)^2  - 1\right)}{16  (\alpha_h +\alpha_y)}  ~{\rm and} ~ \alpha_y\equiv 2\alpha_v /N  \ .
\end{equation}
Using the EOM \eqref{EOM} and that $J=2V\mu b^2$, we can express $\frac{\mu}{m}$ in terms of the parameter $\mathcal{J} \equiv 2 J \frac{\alpha_h + \alpha_y}{N}$ as

\begin{equation}
    \frac{\mu}{m}  = \frac{1}{2}\frac{3^\frac{1}{3}+ x^\frac{2}{3}}{3^\frac{2}{3} x^\frac{1}{3}} \ ,
\end{equation}
where $x= 36 \mathcal{J} + \sqrt{-3+1296 \mathcal{J}^2 }$. The coupling $\mathcal{J}$ controls the transition between the large charge ($\mathcal{J} \gg 1$) and the perturbative ($\mathcal{J} \ll 1$) regimes. Our final expression for the leading contribution to the anomalous dimension $\Delta_{LO}= R E_{LO} $ is
\begin{equation} \label{class2}
    \Delta_{LO} =\frac{N}{144 (\alpha_h +\alpha_y)}\left[ \frac{3^\frac{1}{3}\left(3^\frac{1}{3}+x^\frac{2}{3}\right)^4 }{x^\frac{4}{3}}- 2\frac{3^\frac{2}{3}\left(3^\frac{1}{3}+x^\frac{2}{3}\right)^2}{x^\frac{2}{3}} - 9 \right] 
\end{equation}
and,  notice that in $\Delta_{LO}$ we used generic values for the couplings $\alpha_h $ and $\alpha_y$ since at LO Lagrangian \eqref{model} is Weyl invariant for any values of the couplings.

We now proceed with the computation of the leading quantum corrections $\Delta_{NLO}$. Its bare expression is given by the fluctuation functional determinant and reads
\begin{equation}
    \Delta_{NLO}^{\text{bare}} =  \frac{R}{2}\sum_{\ell=0}^\infty n_{\ell}\left[ \sum_i g_i(N) \omega_i(\ell) \right]\
\end{equation}
where $n_\ell=\frac{(2\ell+d-2) \Gamma(\ell+d-2)}{\Gamma(\ell+1) \Gamma(d-1)}$ is the Laplacian multiplicity on $S^{d-1}$. The inner sum runs over all the dispersion relations $\omega_i$ computed in \eqref{disp}, each counted with its multiplicity $g_i(N)$.

After renormalization, we can express $\Delta_{NLO}$ in terms of a convergent sum which can be computed numerically. Thus, following the procedure of \cite{Badel:2019oxl}, we arrive at our final expression for the NLO contribution in the semiclassical expansion at the fixed points, which reads
\begin{equation} \label{NLO}
    \Delta_{NLO} = \rho  + \frac{1}{2} \sum_{\ell=0}^\infty \left[ R (1+ \ell)^2
\left(\sum_i g_i(N) \omega_i(\ell)\right)_{d=4} + \sigma \right] \ .
\end{equation}
The functions $\rho (\mathcal{J}^*, N, \alpha_h^*,  \alpha_y^*)$ and $\sigma (\ell,\mathcal{J}^*, N, \alpha_h^*,  \alpha_y^*)$ are given in appendix. Our results \eqref{class2} and \eqref{NLO} resum to all-orders in the coupling $\mathcal{J}$ the LO and NLO terms in the charge expansion respectively.
We now focus on the perturbative regime at small $\mathcal{J}$, where the sum of \eqref{class2} and \eqref{NLO} evaluated at the FPs reads
\begin{eqnarray}
\Delta_{{LO}} &+& \Delta_{NLO}  =  Q\left[1+ \frac{ Q (\alpha_h^* +\alpha_y^*) }{N}
\right. \\ &-& \left.  \frac{2(7+ 2 N){\alpha_h^*}^2+(9+ 8 N)\alpha_h^* \alpha_y^* + (5+N^2){\alpha_y^*}^2}{N(\alpha_y^*+\alpha_h^*)} + \mathcal{O}(\mathcal{J}^2)   \right] \nonumber
\end{eqnarray}
where $Q= 4 J$ is the classical dimension of the operator and for $\mathcal{O}(\mathcal{J})$ term we substituted $\mathcal{J}=Q \frac{\alpha_h^* + \alpha_y^*}{2N}$ explicitly.  We checked that for $\alpha_h^* = 0$ the above results reproduce the anomalous dimension of the $Q$-index traceless symmetric $O(2 N^2)$ tensor with classical dimension $Q$. The presence of the couplings at the denominator in the perturbative expansion is a somewhat surprising feature of our results which, at first sight, can look suspicious. Nevertheless, one has to remind oneself that the above expression is strictly valid \emph{only} at the fixed points so one should look at the conformal dimension $\Delta$ as a function of $\epsilon$ and not as a function of the couplings. 

Considering the FPs \eqref{FPS}, we obtain the scaling dimension at $\mathcal{O}(\epsilon)$, which reads
\begin{widetext}
\begin{equation}
    \Delta = Q +\left( \frac{-126 + 10 N + 34 N^2 - 2N^3 - 4 N^4 \pm i\,(6 - 2 N) \sqrt{2 N^2-6}   }{8 (27 - 8 N^2+ N^4)}Q +  \frac{ 18 - 5 N - 2 N^2 + N^3 \pm i\,( 2N -3)  \sqrt{2 N^2-6} }{8 (27 - 8 N^2+ N^4)}Q^2 \right) \epsilon +\mathcal{O}(\epsilon^2) \ .\label{Scaling_Dimension}
\end{equation}
\end{widetext}
From the group-theoretical arguments given in the previous section, 
this result should correspond to the 1-loop scaling dimensions for the family of the charged operators $\mathcal{O}_J=\Tr[(\be_{21} H\be_{12} H^\dagger)^{Q/2}]$ . 
 {Remarkably, for $Q=2$ and at 1-loop this result (Eq.~\eqref{Scaling_Dimension})  matches the anomalous dimension of the bi-adjoint operator $\rm{Tr}\left[T^a H T^a H^{\dagger} \right]$ at the fixed point Eq.~\eqref{FPS} shown in  table \eqref{table}.}
We can combine our semiclassical results with the knowledge of the $2$-loop anomalous dimension for $\rm{Tr}\left[T^a H T^a H^{\dagger} \right]$ \cite{Antipin:2014mga}, to extract the complete $2$-loop anomalous dimension for the whole family. Remarkably, assuming  ordinary perturbative power series expansion in both couplings combined with the constraint that for $\alpha_h=0$ we reproduce the known result for the $O(2N^2)$ model \cite{Antipin:2020abu}, we can 
write it in a form valid  beyond the FPs, i.e it holds for any perturbative values of the couplings. We have
\begin{widetext}
\begin{align} \label{twoloop}
    \Delta_{\text{2-loops}} = & Q+\frac{Q (Q-1) \alpha_y}{N} +\frac{Q (Q-2) \alpha_h}{N} - \textcolor{red}{Q \left[2 \left(\frac{3}{N^2}- \frac{4}{N} - 1\right)\alpha_h^2 + 4\left(\frac{2}{N^2}-\frac{3}{N} \right) \alpha_h \alpha_y + \frac{1}{2}\left(\frac{1}{N^2}- 3\right)\alpha_y^2\right]} \nonumber \\ & +Q^2 \left[2 \left(\frac{1}{N^2}-\frac{2}{N} \right)\alpha_h^2 + 4 \left( \frac{3}{N^2} - \frac{2}{N}\right) \alpha_h \alpha_y + \left(\frac{3}{N^2} - 1 \right)\alpha_y^2  \right] - \frac{2 Q^3 (\alpha_h +\alpha_y)^2}{N^2} \ ,
    \end{align}
\end{widetext}
where we highlighted in red the term that was not predicted by our semiclassical result as it is an NNLO in the charge expansion.
 It would be interesting to extend this strategy to higher loops. 
 
The above shows that the semiclassical approach can be successfully applied to near-conformal physical theories featuring complex CFTs.   

Before concluding we note that the charge configuration used so far differs from the one adopted in 
 \cite{Orlando:2019hte} when investigating the large charge regime of the $U(N)\times U(N)$ linear sigma model embedded in a safe theory. In their case the charge assignment reads $\mathcal{Q}_{L,J}^*=-\mathcal{Q}_{R,J}^*=\rm{diag}\{\underbrace{J,\cdots,J}_{N/2},\underbrace{-J,\cdots,-J}_{N/2}\}$.
For such configuration, the classical dimension $Q$ of the corresponding fixed-charge operator satisfies $ Q= 2 J N$.  As discussed earlier $J$ must be positive and integer or half-integer with the minimal value 1/2,  thus in this case $Q$ is non-trivially related to $N$ by the constraint
$Q \ge N$. 
As a consequence, for a given $Q$, the irreducible representation to which the operator
belongs depends on the value of $N$, making the identification of such operators highly non-trivial. 
For instance, using the semiclassical method exploited here we have checked that for $Q=2$ and arbitrary $N > 2$ the corresponding result does not match any of the operators in table \eqref{table}, while for $N=2$ and arbitrary $Q$ it coincides with Eq.\eqref{twoloop}. Notice that this is a simple consequence of the fact that for $N = 2$ this charge configuration coincides with ours. The bottom line is that one needs to choose which charge configuration to consider with care when analyzing the fixed charge sectors of a CFT. We will further analyze this issue and related ones in a follow-up paper.

\medskip
\section*{Acknowledgement}
The work of O.A. and J.B. is partially supported by the Croatian Science Foundation project number 4418 as well as European Union through the European Regional Development Fund - the Competitiveness and Cohesion Operational Programme (KK.01.1.1.06). F.S and Z.W acknowledge the partial support by Danish National Research Foundation grant DNRF:90.

\appendix

\section{A ~ ~ ~ ~Operator Identification from Charge Configurations}

For simplicity, we confine ourselves to charge configurations that satisfy $\mathcal{Q}_L+\mathcal{Q}_R=0$,
dictated by the diagonal ansatz solution, and simply write $\mathcal{Q}_L$ as $\mathcal{Q}$. $\mathcal{Q}$
lives in the Cartan subalgebra $\mathcal{H}$ of $SU(N)$, and its meaning is characterized by the set of charge eigenvalues
associated with a set of orthonormal basis elements in $\mathcal{H}$. Suppose $\hat{h}$ is one of the basis
elements. In the self-representation of $SU(N)$ in which $\hat{h}$ is a traceless diagonal $N\times N$ matrix, we determine
the proper normalization condition for $\hat{h}$ as $\Tr(\hat{h}^2)=\frac{1}{2}$. For example $\hat{h}_j\equiv
\frac{1}{2}(\be_{j,j}-\be_{j+1,j+1})$ for $j=1,2,...,N-1$ are normalized elements (although one should be careful that for adjacent
$j$'s the elements are not orthogonal). For any normalized basis element $\hat{h}$, the associated charge eigenvalue $q_h$
for a given charge configuration $\mathcal{Q}$ can be computed as
\begin{align}
q_h=2\Tr(\mathcal{Q}h) \ .
\end{align}
Note only for orthonormal basis elements this is equivalent to the coefficient extracted from the decomposition of $\mathcal{Q}$
onto basis elements, and different choices of orthonormal basis give charge eigenvalues that are compatible with each other. The key relation in fixing all these normalizations is the following commutation relation in $SU(N)$ Lie algebra:
\begin{align}
[\hat{h}_j,\be_{pq}]=\frac{1}{2}(\delta_{jp}-\delta_{jq}-\delta_{j+1,p}+\delta_{j+1,q})\be_{pq}
\end{align}
for $j=1,2,...,N-1$ and $p,q=1,2,...,N$ with $p\neq q$. A special case is
\begin{align}
[\hat{h}_j,\be_{j,j+1}]=\be_{j,j+1}
\end{align}
when this equation is turned into a commutation relation between Noether charge and fixed-charge operators
constructed from fields that satisfy canonical commutation relations. The standard normalization condition
introduced above is then implied.

The general method to construct fixed-charge operator with the minimal classical scaling dimension corresponding to a given charge configuration
starts with building blocks that have simple definite transformation properties under $SU(N)_L\times SU(N)_R\times U(1)_A$.
Since we are concerned with charge configurations that satisfy $\mathcal{Q}_L+\mathcal{Q}_R=0$, the building block
takes the form $\Tr(\tau H\tau^T H^\dagger)$ with $\tau$ being an element in some root subspace of the $SU(N)$ Lie algebra.
Obviously, this object lives in the bi-adjoint representation of $SU(N)_L\times SU(N)_R$. To build operators with
more fields one replicates the same structure inside the same trace operation, such as
\begin{align}
\Tr\Big[ \Pi_j (\tau_j H\tau_j^T H^\dagger)^{|y_j|}\Big] \ .
\label{eq:ocon}
\end{align}
Here $y_j$ is an integer, and $\tau_j=\be_{pq}$ for some $p,q$ that depend on $j$.
More generally, one may choose to redistribute the trace operation (i.e. splitting one single trace to multiple traces),
and changing the order of matrix products for different $\tau_j H\tau_j^T H^\dagger$ factors, to obtain more operators associated with the same charge configuration. The value of $y_j$ should be determined from the charge configuration $\mathcal{Q}$. Since
we consider operators with the minimal classical scaling dimension, it suffices to decompose $\mathcal{Q}$
onto a root basis ${\beta_j},j=1,2,...,N-1$, which in precise terms mean to consider the decomposition
\begin{align}
\mathcal{Q}=\Sigma_{j=1}^{j=N-1}y_j \hat{h}_{\beta_j}
\label{eq:qdec}
\end{align}
where $\hat{h}_{\beta_j}$'s are the roots mapped into $\mathcal{H}$ which satisfy the standard normalization
condition $\Tr(\hat{h}_{\beta_j}^2)=\frac{1}{2}$. Note $\hat{h}_{\beta_j}$'s are not orthogonal, and
$y_j$'s do not correspond to charge eigenvalues. However, by making connection to an arbitrary orthonormal basis,
one may prove that when $\tau_j$ in Eq.~\eqref{eq:ocon} is chosen to be an element in the root subspace of $\beta_j$,
$y_j$ should be determined from the decomposition in Eq.~\eqref{eq:qdec}.

For instance, for a class of charge configurations $\mathcal{Q}_G=\rm{diag}\{Q_1,-Q_1,Q_3,-Q_3,\cdots,Q_{N-1},-Q_{N-1}\}$,
if we choose the root basis such that $\hat{h}_{\beta_j}=\hat{h}_j$, we find the nonzero $y_j$'s are given by
$y_1=2Q_1,y_3=2Q_3,...,y_{N-1}=2Q_{N-1}$. Since $y_j$'s must be an integer, this implies $Q_1,Q_3,...,Q_{N-1}$
must be integers or half-integers. This is not surprising since our charge configuration corresponds to a weight
of Lie algebra representation and thus can only take a discrete set of values.

The number of operators that can be constructed in this manner grows drastically if we realize that we have
the freedom to change the root basis and redo the decomposition, to redistribute the trace operation and
to utilize noncommutativity of matrix products. In general, we need to change the root basis in all possible manners
to find one or more optimal root basis that leads to the minimal classical scaling dimension (i.e. minimizing
the sum of $|y_j|$). After the optimal root bases are found, we need to consider the above-mentioned variations
that all correspond to the same charge configurations. Algebraically some of the variations may be identical.
Moreover in general all the operators constructed in this manner only correspond to the same charge configuration, or
weight, however it is not guaranteed that they are already organized into definite irreducible representations.

Due to these algebraic complications, a complete operator identification for a generic charge configuration looks
quite difficult. Nevertheless, we found a special family of charge configurations defined by $\mathcal{Q}_{L,J}=-\mathcal{Q}_{R,J}=\rm{diag}\left\{-J,J,0,\cdots,0\right\}$
lead to great simplification since one can prove that it corresponds to a unique fixed-charge operator
$\mathcal{O}_J=\Tr[(\be_{21} H\be_{12} H^\dagger)^{2J}]$ living in the representation $(\Gamma_J,\Gamma_J)$ of $SU(N)_L\times SU(N)_R$, with the irreducible representation $\Gamma_J$ of $SU(N)$ defined through
its Dynkin label $(2J,0,\cdots,0,2J)$. Any variations either lead to operators with larger classical scaling dimensions,
or lead to the same operator written in a different form. The uniqueness is related to the fact that this
special family of charge configurations corresponds to highest weights in tensor product of the adjoint representations.

On the other hand, if we consider $\mathcal{Q}_F=\rm{diag}\{J,-J,J,-J,\cdots,J,-J\}$ ($J>0$) which is the charge configuration
used in ~\cite{Orlando:2019hte}, then it does not correspond to a unique operator in general. Moreover,  according to the decomposition in Eq.~\eqref{eq:qdec}, it should correspond to
operators with minimal classical scaling dimension $Q=2NJ$.
Since the minimal nonzero value of $J$ is $\frac{1}{2}$, it implies for this family of charge configurations $Q\geq N$.
Only for $N=2,J=\frac{1}{2}$, $\mathcal{Q}_F$ reduces to $\mathcal{Q}_{1/2}$ which corresponds to the bi-adjoint of
$SU(2)_L\times SU(2)_R$.

\section{B ~ ~ ~ ~The functions $\rho (\mathcal{J}^*, N, \alpha_h^*, \alpha_y^*)$ and $\sigma (\ell, \mathcal{J}^* , N,  \alpha_h^*, \alpha_y^*)$}

Here we provide explicit expressions for the functions $\rho (\mathcal{J}^*, N, \alpha_h^*, \alpha_y^*)$ and $\sigma (\ell, \mathcal{J}^* , N,  \alpha_h^*, \alpha_y^*)$, which appear in our result \eqref{NLO} for the NLO contribution to the anomalous dimension in the semiclassical expansion.

Recalling that $\alpha_y^*= \frac{2\alpha_v^*}{N} $ and that $x^*= 36 \mathcal{J^*} + \sqrt{-3+1296 {\mathcal{J}^*}^2 }$, we have
 \begin{widetext}
\begin{eqnarray}
 && \rho (\mathcal{J}^*, N, \alpha_h^*, \alpha_y^*)  = \frac{1}{240 \left(\alpha_h^*+\alpha _y^*\right)^2} \left[ \frac{10}{3} {\alpha _h^*}^2 \left(-72 N^2-\frac{32 \ \ 3^{1/3} N
   \left({x^*}^{2/3}+ 3^{1/3}\right)^4}{{x^*}^{4/3}}-\frac{48\ 3^{2/3} N \left({x^*}^{2/3}+ 3^{1/3}\right)^2}{{x^*}^{2/3}}+162 N \right.  \right. \nonumber \\ && \left. -\frac{\left(4\ 3^{2/3} {x^*}^{4/3}+15 {x^*}^{2/3}+12   \ 3^{1/3}\right) \left(12  \ 3^{1/3} \alpha_h^* \left(2 \alpha_h^*+\alpha_y^*\right)+4\ 3^{2/3} {x^*}^{4/3} \alpha_h^* \left(2
   \alpha_h^*+\alpha_y^*\right)+3 {x^*}^{2/3} \left(18 {\alpha_h^*}^2+5 \alpha_h^* \alpha_y^*-4 {\alpha_y^*}^2\right)\right)}{{x^*}^{4/3} \left(\alpha_h^*+\alpha_y^*\right)^2}  \right)  \nonumber \\ && +\frac{20}{3} \alpha_h^* \alpha_y^* \left(-\frac{\left(4\ 3^{2/3} {x^*}^{4/3}+15
   {x^*}^{2/3}+12   \  3^{1/3}\right) \left(12  \  3^{1/3} \alpha_h^* \left(2 \alpha_h^*+\alpha_y^*\right)+4\ 3^{2/3} {x^*}^{4/3} \alpha_h^* \left(2 \alpha_h^*+\alpha_y^*\right)+{x^*}^{2/3} \left(62 {\alpha_h^*}^2+31 \alpha_h^* \alpha_y^*-4 {\alpha_y^*}^2\right)\right)}{{x^*}^{4/3} \left(\alpha_h^*+\alpha_y^*\right)^2} \right.  \nonumber \\ && \left. -\frac{8\ 3^{2/3} N^2 \left({x^*}^{2/3}+ 3^{1/3}\right)^2}{{x^*}^{2/3}}-54 N^2-\frac{32   \  3^{1/3}N
   \left({x^*}^{2/3}+ 3^{1/3}\right)^4}{{x^*}^{4/3}}-\frac{16\ 3^{2/3} N \left({x^*}^{2/3}+  3^{1/3}\right)^2}{{x^*}^{2/3}}+90 N \right) + {\alpha_y^*}^2 \left(10 \left(\frac{4 \left({x^*}^{2/3}+ 3^{1/3}\right)^2}{3 \
    3^{1/3} {x^*}^{2/3}}-1\right)  \right.  \nonumber \\ && \left. \left. \left(-\frac{45 \alpha_h^*}{\alpha_h^*+\alpha_y^*}-\frac{9 {\alpha_h^*}^2}{\left(\alpha_h^*+\alpha_y^*\right)^2}-\frac{4 \ 3^{2/3} \left({x^*}^{2/3}+ 3^{1/3}\right)^2 \left(8
   {\alpha_h^*}^2+7 \alpha_h^* \alpha_y^*+2 {\alpha_y^*}^2\right)}{{x^*}^{2/3} \left(\alpha_h^*+\alpha_y^*\right)^2}-18\right)+N^2 \left(-\frac{40 \left(2 {x^*}^{8/3}+33 {x^*}^{2/3}+11 \  3^{1/3} {x^*}^2+6 \
   3^{1/3}\right)}{3^{2/3} {x^*}^{4/3}}-825\right)\right)  \right] \nonumber \\
\end{eqnarray}

and
 
\begin{eqnarray}
 && \sigma (\ell, \mathcal{J}^*, N, \alpha_h^*, \alpha_y^*)  = \frac{1}{4 \ell \left(\alpha_h^*+\alpha _y^*\right)^2} \left[ 2 {\alpha_h^*}^2 \left( -4 \ell (\ell+1)^3 N^2-2 N \left(\frac{4
   \left({x^*}^{2/3}+3^{1/3}\right)^2}{3 \  3^{1/3} {x^*}^{2/3}}-1\right) \left((2 \ell+1)^2-\frac{4 \left({x^*}^{2/3}+3^{1/3}\right)^2}{3 \ 3^{1/3} {x^*}^{2/3}}\right)+ \left(4\ 3^{2/3} {x^*}^{4/3}+15 {x^*}^{2/3}\right.  \right.  \right. \nonumber \\ && \left. \  +12 \ 3^{1/3}\right)  \left.  \frac{\alpha_h^* \alpha_y^* \left(3 \left(6 \ell^2+6 \ell+5\right) {x^*}^{2/3}+4\ 3^{2/3} {x^*}^{4/3}+12 \  3^{1/3}\right) +\left(8\
   3^{2/3} {x^*}^{4/3}+30 {x^*}^{2/3}+24 \  3^{1/3}\right) {\alpha_h^*}^2+18 \ell (\ell+1) {x^*}^{2/3} {\alpha_y^*}^2}{27 {x^*}^{4/3} \left(\alpha_h^*+\alpha_y^*\right)^2}\right) \nonumber \\ && +4 \alpha_h^* \alpha_y^*
   \left(-\ell (\ell+1) N^2 \left(4 \ell
   (\ell+2)+\frac{4 \left({x^*}^{2/3}+3^{1/3}\right)^2}{3 \ 3^{1/3} {x^*}^{2/3}}+3\right)-2 N \left(\frac{4 \left({x^*}^{2/3}+3^{1/3}\right)^2}{3 \ 3^{1/3} {x^*}^{2/3}}-1\right) \left(2 \ell (\ell+1)-\frac{4
   \left({x^*}^{2/3}+3^{1/3}\right)^2}{3 \ 3^{1/3} {x^*}^{2/3}}+1\right)+\left(4\ 3^{2/3} {x^*}^{4/3}+15 {x^*}^{2/3} \right. \right.  \nonumber
    \end{eqnarray}
\begin{eqnarray}
 && \left. \left.+12 \ 3^{1/3}\right)  \frac{\alpha_h^* \alpha_y^* \left(-3 \left(2 \ell^2+2 \ell-5\right) {x^*}^{2/3}+4\ 3^{2/3} {x^*}^{4/3}+12 \ 3^{1/3}\right) +2 {\alpha_h^*}^2
   \left(-3 \left(2 \ell^2+2 \ell-5\right) {x^*}^{2/3}+4\ 3^{2/3} {x^*}^{4/3}+12 \ 3^{1/3}\right)+6 \ell (\ell+1) {x^*}^{2/3} {\alpha_y^*}^2 }{{27 {x^*}^{4/3} \left(\alpha_h^*+\alpha_y^*\right)^2}} \right)  \nonumber \\ &&  + {\alpha_y^*}^2 \left(2 \left(\frac{4 \left({x^*}^{2/3}+3^{1/3}\right)^2}{3 \ 3^{1/3} {x^*}^{2/3}}-1\right) \left(-3 \alpha_h^*\frac{2 \ell^2
   + 2 \ell  +1 }{\alpha_h^*+\alpha_y^*}-\frac{3 {\alpha_h^*}^2}{\left(\alpha_h^*+\alpha_y^*\right)^2}+\frac{4
   \left({x^*}^{2/3}+3^{1/3}\right)^2 \left(8 {\alpha_h^*}^2+7 \alpha_h^* \alpha_y^*+2 {\alpha_y^*}^2\right)}{3 \ 3^{1/3} {x^*}^{2/3} \left(\alpha_h^*+\alpha_y^*\right)^2}-2 \ell^2-2 \ell-2\right) \right. \nonumber \\ && \left. \left. - N^2 \left(4
   \left(\frac{4 \ell (\ell+1) \left({x^*}^{2/3}+3^{1/3}\right)^2}{3 \ 3^{1/3} {x^*}^{2/3}}+\ell (\ell+1) (2 \ell (\ell+2)+1)-\frac{4 \left({x^*}^{2/3}+3^{1/3}\right)^4}{9\ 3^{2/3} {x^*}^{4/3}}\right)+\frac{8
   \left({x^*}^{2/3}+3^{1/3}\right)^2}{3 \ 3^{1/3} {x^*}^{2/3}}-1\right)\right)   \right]  \ \ .
\end{eqnarray}
\end{widetext}



  \end{document}